# The Origin of the Pseudogap from Tunneling Spectroscopy Measurements on $Bi_2Sr_2CaCu_2O_{8+x}$ Single Crystals.

A. Mourachkine

*Université Libre de Bruxelles, Service Physique des Solides, CP233, Boulevard du Triomphe, B-1050 Brussels, Belgium*

We present electron-tunneling spectroscopy of oxygen over- and under-doped $Bi_2Sr_2CaCu_2O_{8+x}$ single crystals in a temperature range between 14 K and 290 K using a break-junction technique. In over- and under-doped samples we observe a pseudogap with $T^* = 280 - 290$ K and $T^* > 290$ K, respectively. In order to explain the data we consider possible models for the pseudogap, namely, (i) a precursor pairing; (ii) a charge-density-wave gap; (iii) a spin-gap due to antiferromagnetic (AF) correlations. Analyzing all data we show unambiguously that the only model which fits the data is a gap due to AF correlations.

The existence of a pseudogap in electronic excitation spectra of high-$T_c$ superconductors (HTSC), which appears below a certain temperature $T^* > T_c$, is considered to be amongst the most important features of cuprates. Many experiments have provided evidence (NMR [1], angle-resolved photoemission (ARPES) [2], specific heat [3], electron-tunneling spectroscopy (ETS) [4] and STM [5]) for a gaplike structure in electronic excitation spectra. In all these studies, the pseudogap (PS gap) was reported to exist in underdoped HTSC. However, ETS [4] and STM [5] measurements showed that the PS gap is present in overdoped samples as well. Indeed, the tunneling spectroscopy is particularly sensitive to the density of state (DOS) near the Fermi level $E_F$ and, thus, is capable of detecting *any* gap in the quasiparticle excitation spectrum at $E_F$ [6]. In addition to this, it has a very high energy resolution (~ $k_BT$) [6]. The tunneling spectroscopy performed by STM and a break-junction (B-J) technique has an additional advantage: to measure the density of state (DOS) *locally*. From earlier studies, the PS gap was associated with a spin-gap in the antiferromagnetic (AF) excitation spectrum [1] or with a spin-density-wave (SDW) gap due to AF correlations [7]. However, the interpretation of the recent ARPES measurements [8] supports the view that the PS gap is a precursor pairing rather than the magnetic origin. At the same time, some NMR and ARPES data for the PS gap [9] have been interpreted as a charge-density-wave (CDW) gap [10]. This raises the question about the origin of the PS gap.

It is well known that the parent compounds of the HTSC are AF insulators. The tendency of an antiferromagnet to expel holes leads to the formation of the hole-rich and hole-free regions in cuprates [11]. When holes are doped into the $CuO_2$ planes, AF correlations survive and coexist with superconductivity [12].

In this Letter, we present direct measurements of the density of states by electron-tunneling spectroscopy on over- and under-doped $Bi_2Sr_2CaCu_2O_{8+x}$ (Bi2212) single crystals using a break-junction technique. We detect a pseudogap in over- and under-doped samples with $T^* = 280 - 290$

K and $T^* > 290$ K, respectively. To our knowledge, it is the first time the measurement of a PS gap on an overdoped Bi2212 with $T^* = 280$ K is presented in the literature. In order to fit the data we consider all proposals for the model of the PS gap: (i) the precursor pairing; (ii) the CDW gap, and (iii) the spin-gap due to AF correlations. We find that the PS gap in our study can only be associated with the gap due to AF correlations.

The single crystals of Bi2212 were grown using a self-flux method and then mechanically separated from the flux in $Al_2O_3$ or $ZrO_2$ crucibles [13]. The dimensions of the samples are typically $3\times1\times0.1$ mm$^3$. The chemical composition of the Bi-2212 phase corresponds to the formula $Bi_2Sr_{1.9}CaCu_{1.8}O_{8+x}$ in overdoped crystals as measured by energy dispersive X-ray fluorescence (EDAX). The crystallographic *a-*, *b-*, *c-*values of the overdoped single crystals are of 5.41 Å, 5.50 Å and 30.81 Å, respectively. The $T_c$ value was determined by either dc-magnetization or by four-contacts method yielding $T_c = 87 - 90$ K with the transition width $\Delta T_c \sim 1$ K. Underdoped samples were obtained from the overdoped single crystals by annealing the crystals in vacuum [5]. The underdoped samples which were studied have the critical temperatures of 21 K, 51 K and 76 K. Experimental details of our B-J technique are given in Refs. [13, 14].

Typical conductance curves *dI/dV(V)* and tunneling current-voltage *I(V)* characteristics for our superconductor-insulator-superconductor (SIS) break junctions on Bi2212 single crystals can be found elsewhere [14]. They exhibit the characteristic features of typical SIS junctions [15, 16]. The magnitude of a superconducting (SC) gap can, in fact, be derived directly from the tunneling spectrum. However, in the absence of a generally accepted model for the gap function and the DOS in HTSC, such a quantitative analysis is not straightforward [5]. Thus, in order to compare different spectra, we calculate the gap amplitude $2\Delta$ (in m*e*V) as a half spacing between the conductance peaks at $\pm 2\Delta$.

Most of our study was carried out on slightly overdoped Bi2212 single crystals. Figure 1 shows a set of normalized tunneling conductance curves measured between 14 K and 290 K as a function of bias voltage on an overdoped Bi2212 single crystal with $T_c = 88.5$ K. The value of the SC gap at 14 K is $2\Delta = 45$ m*e*V. All curves except the last one show a gaplike structure. There is no sign indicating at what temperature the SC gap was closed. Across $T_c$, the SC tunneling spectra evolve continuously into a normal state quasiparticle gap structure which vanishes at 232 K < $T^*$ 290 K but remains almost unchanged with temperature up to 232 K. These data are similar to data obtained on an underdoped Bi2212 single crystal by STM [5]. In Fig. 2 we show another set of normalized tunneling conductance curves measured between 14 K and 290 K (shown only between 14 K and 104 K [17]) as a function of voltage on another overdoped sample with $T_c = 89.5$ K. The shape of the conductance curve at 14 K is slightly different from the similar conductance curve in Fig. 1 but the value of the SC gap is almost the same, $2\Delta = 46$ m*e*V. It can be seen in Fig. 2 that the conductance curves at 78.5 K and at 80 K have two gaplike structures. A closer inspection of the conductance curves between 81 K and 90 K shows that the smaller gap disappears quickly with increase of temperature while the larger one increases with $T$. Across $T_c$, this gap-like structure evolves into a slightly asymmetrical PS gap with $T^* = 285$ K [17]. Most spectra for overdoped samples ($T_c = 87 - 90$ K) in our study show $T^* = 280 - 290$ K.

We now discuss the temperature dependence of the SC gap and the PS gap in overdoped Bi2212. It is very difficult to derive the value of a PS gap from tunneling spectra shown in Fig. 1 at high temperatures. Therefore, in Fig. 3 we present a temperature dependence of the SC gap and the PS gap only for temperatures between 0 $\leq T/T_c \leq$ 1.2 [18]. The curve A in Fig. 3 corresponds to the temperature dependence of the tunneling spectra shown in Fig. 1. The curves B and C depict the temperature dependence of the tunneling spectra shown in Fig. 2. The curve B in Fig. 3 corresponding to the $\Delta(T)$ is clearly below the BCS-dependence. It was shown earlier that in two-band superconductivity, the temperature dependence of the smaller order parameter (OP) [19] appears below the BCS-temperature dependence [6, 20]. This means that in overdoped Bi2212 there are two OPs. The curve C in Fig. 3 increases with $T$ as $T$ approaches $T_c$ and then evolves into a PS gap. It indicates that the second OP which has to be larger than the SC OP corresponds to the PS-gap OP. We can estimate the value of this OP from $T^* \sim \Delta(0)/2$ [11]. Using $T^* = 290$ K we have $\Delta_{PS} \sim 50$ meV. One can see in Fig. 3 that the curve A can be presented by a combination of two different curves: at low temperatures $T/T_c \leq 0.5$ by the curve B shown in Fig. 3 and at higher temperatures $0.6 < T/T_c \leq 1.2$ by almost a flat, temperature-independent curve. It is obvious that the temperature-independent curve corresponds to the PS gap. The first part of the curve A which is below the BCS-temperature dependence, like the curve B, indicates the presence of two OPs: the PS and SC OPs. The presence of the Josephson current in the spectra shown in Fig. 1 at high temperatures means that the PS gap is also a SC gap or that the SC-gap peaks are below the PS-gap peaks. The results which are presented in Fig. 3 are not unique. Our measurements show a few $\Delta(T)$ curves similar to the curve A and a few ones like curve B. These different curves are obtained in different overdoped samples. So, we believe that our results reflect the intrinsic properties of overdoped Bi2212 single crystals.

We now turn to the temperature dependence of the quasiparticle DOS in an underdoped and an optimally doped Bi2212. At the moment we present only the temperature dependence of the OP in an underdoped Bi2212. More detailed information of our study on underdoped Bi2212 single crystals will be presented elsewhere. The inset in Fig. 3 shows a temperature dependence of the $\Delta(T)$ for an underdoped Bi2212 single crystal with $T_c = 51$ K. The data are in a good agreement with data obtained by STM [21]. It is clear that the temperature dependence of the OP for the underdoped Bi2212 is well above the BCS-temperature dependence. The $T^*$ value lies above 290 K. In the inset of Fig. 3 we also present data for an optimally doped Bi2212 single crystal taken from Ref. [16]. One can see that the total $\Delta(T)$ for the optimally doped Bi2212 has some tendency to lie below the BCS-temperature dependence as in overdoped samples.

In order to understand the relation between the SC gap and the PS gap for over- and under-doped Bi2212 single crystals in Fig. 4 we present the maximum values of the SC and PS gaps at 14 K as a function of the hole concentration, $p$. The hole concentration has been obtained from the empirical relation $T_c/T_{c,\,max} = 1 - 82.6(p - 0.16)^2$ which is satisfied for a number of HTSC [9] and we use $T_{c,\,max} = 95$ K. The requirements for the SC gap to distinguish it from the PS gap were the shape of the conductance curve and the presence of the Josephson current [22]. The PS gap below $T_c$ has a different shape from the SC gap [23]. Measurements have been performed on underdoped Bi2212 single crystals with $T_c$ = 21 K, 51 K, and 76 K and on overdoped samples which were described above. Since at 14 K it was not possible to detect separately a PS gap in overdoped Bi2212 samples, we decided to take a value estimated above $\Delta_{PS}$ = 50 meV. The errors of the measured data in Fig. 4 for the SC and PS gaps at 14 K are small ~ $2(k_B T)$ = 2.4 meV [6]. The points (0.05, 0) and (0.27, 0) for the SC gap are obvious from the fact that $T_c = 0$. For the SC-gap values, we find a good agreement between 0.11 and 0.19 with tunneling data also obtained by the B-J technique [16] and shown in Fig. 4. Measured values of the PS-gap shown in Fig. 4 are maximum in our study but in reality may be larger. Now we are ready to explain the difference in the temperature dependencies of the OP for over- and under-doped Bi2212 single crystals. The difference between two gaps at low hole concentration is large. The SC gap which is inside of the PS gap diminishes with increase of temperature while the PS gap remains almost unchanged. It creates an effect of increasing gap with increase of $T$. For overdoped Bi2212 the ratio $\Delta_{PS,\,max}/\Delta_{SC,\,max}$ becomes smaller. The interaction between two OPs enhances. Two OPs compete with each other. It seems that the PS gap is dominant. In fact, the shape of the $\Delta(T)$ curve is determined not only by the hole concentration but also at what angle on surface of a $d$-wave PS gap the DOS is tested. It is possible that the point $p = 0.16$ for optimally doped Bi2212 is determined by the ratio $\Delta_{PS,\,max}/\Delta_{SC,\,max}$.

Finally, we discuss possible models for the PS gap: (i) a precursor pairing; (ii) a CDW gap, and (iii) a spin-gap due to AF correlations. (i) From the discussion above it is clear that the SC gap and the PS gap are two *different* gaps. The SC gap develops on top of the PS gap. (ii) A CDW gap is a good candidate for the PS gap. In Fig. 4 we show schematically a Balseiro-Falicov (BF) model for a CDW [24, 10], in which a CDW OP competes with a SC OP. At low hole concentration there is a good agreement between the BF model and the measured data for the PS gap. However, it is not the case for high hole concentration. It seems that the PS gap is always larger than the SC gap. Recently it has been shown that the gap amplitude at the $d$-wave nodes scales with $T_c$ [25] but the maximum gap scales with $T^*$ [26]. Our data (see Fig. 4) are completely in agreement with these statements. A "pure" CDW scenario does not fit our data but, nevertheless, a CDW may be present in cuprates. (iii) We could not find any contradiction for the scenario in which the PS gap is a spin-gap due to AF correlations. Moreover, there is a good agreement with the theory [11]. The difference between the SC gap and the spin-gap has a tendency to be "smoothed out" by the motion of the holes between the hole-rich and hole-free regions. This effect is stronger in overdoped Bi2212 where the hole concentration is higher and the difference between the two gaps is smaller. The spin-gap is a basis for the development of the SC and, at the same time, "kills" the SC in overdoped Bi2212. In this scenario in which the PS gap is a spin-gap due to AF correlations, a CDW may be present [11]. It is possible that in this "mixed" AF - CDW scenario the point $p = 0.16$ for optimally doped Bi2212 is determined by a disappearance of the CDW order.

In summary, we presented direct measurements of the density-of-state by tunneling spectroscopy on over- and under-doped $Bi_2Sr_2CaCu_2O_{8+x}$ single crystals in a temperature range between 14 K and 290 K using a break-junction technique. We show that a pseudogap observed in our study can be associated with a spin-gap due to antiferromagnetic correlations, probably, with the presence of a charge-density wave. Other possible models do not fit the data. The $T^*$ obtained in various experiments is to be considered *in most* experiments as a characteristic energy scale and not as a temperature where the pseudogap is reduced to zero.

I thank R. Deltour for support, Y. DeWilde and D. N. Davydov for discussions, H. Hancotte for help with experiment and D. Ciurchea for EDAX examination. This work is supported by PAI 4/10.

_________________________

FIGURE CAPTIONS:

FIG. 1. Tunneling spectra for a SIS junction measured as a function of temperature on an overdoped Bi2212. The conductance scale corresponds to the 290 K spectrum, the other spectra are offset vertically for clarity. The curves have been normalized at -150 mV (or nearest point).

FIG. 2. Tunneling spectra for a SIS junction measured as a function of temperature on an overdoped Bi2212. The conductance scale corresponds to the 104 K spectrum, the other spectra are offset vertically for clarity. The curves have been normalized at -100 mV (or nearest point).

FIG. 3. Measured temperature dependence of the quasiparticle DOS on overdoped Bi2212: the curve A, diamonds (for tunneling spectra shown in Fig. 1); the curve B, triangles and the curve C, dots (for tunneling spectra in Fig. 2). The solid line corresponds to the BCS-temperature dependence. Inset: normalized $\Delta(T)$ vs. $T$: diamonds ($T_c = 51$ K, an underdoped Bi2212) and dots ($T_c = 95$ K, an optimally doped, Ref. [16]). Axis labels in inset coincide with the labels in the main figure. The dashed lines are guides to the eye.

FIG. 4. Maximum gap value at 14 K vs. hole concentration in Bi2212 single crystals: diamonds (PS gap); dots (SC gap, $\Delta_{SC}$), and triangles (SC gap, Ref. [16]). The continuos line corresponds a Balseiro-Falicov model for a CDW [24]. The dashed lines are guides to the eye.

**(NOTE: the figures are in .ps format with poor quality for graphs)**

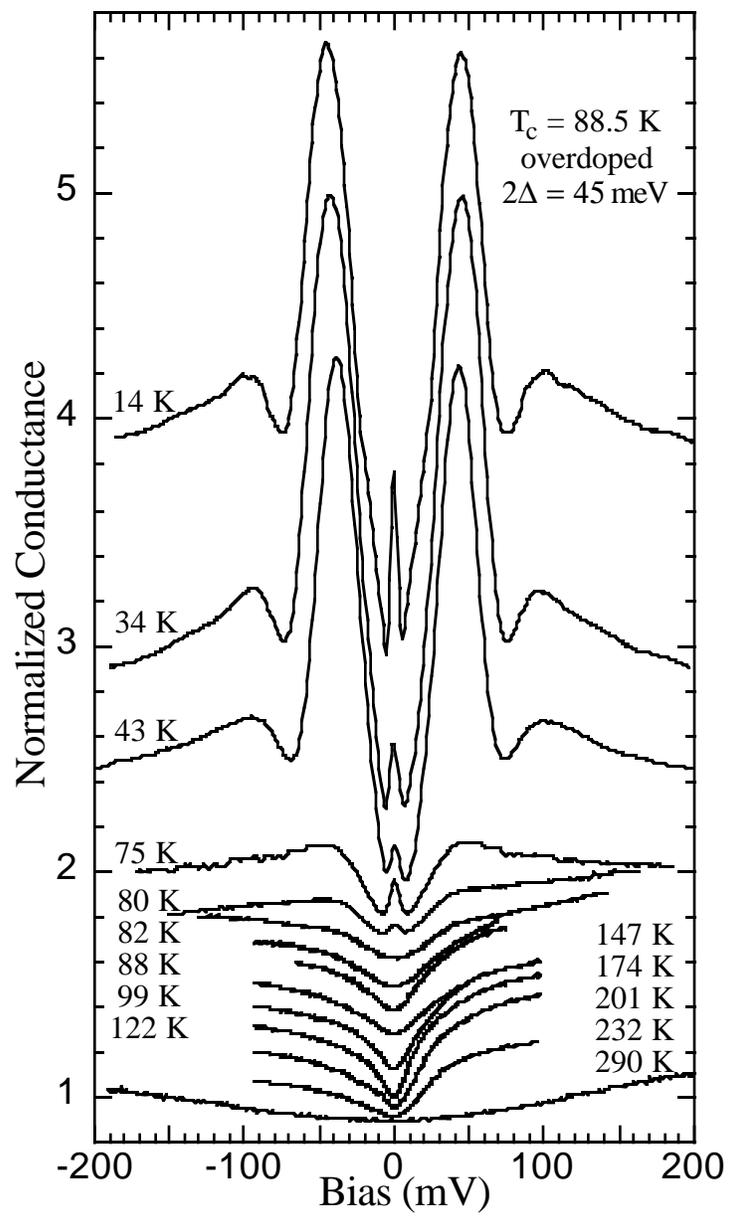

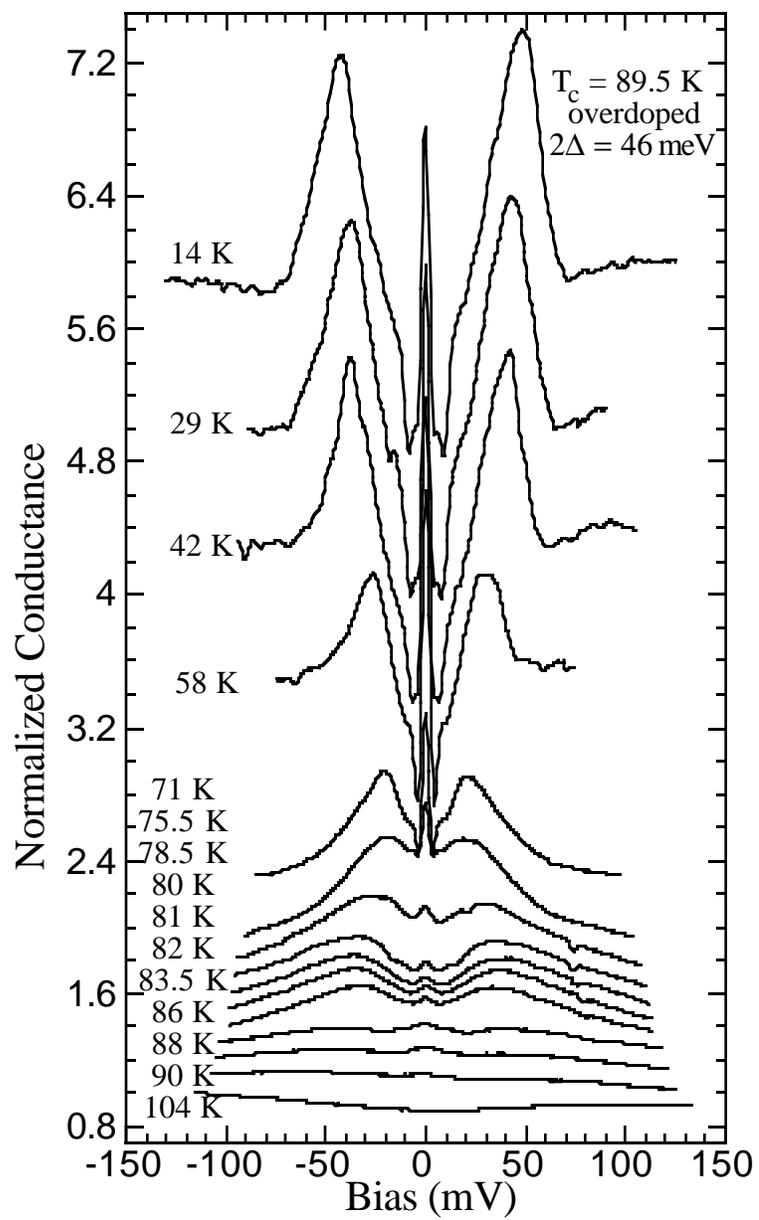

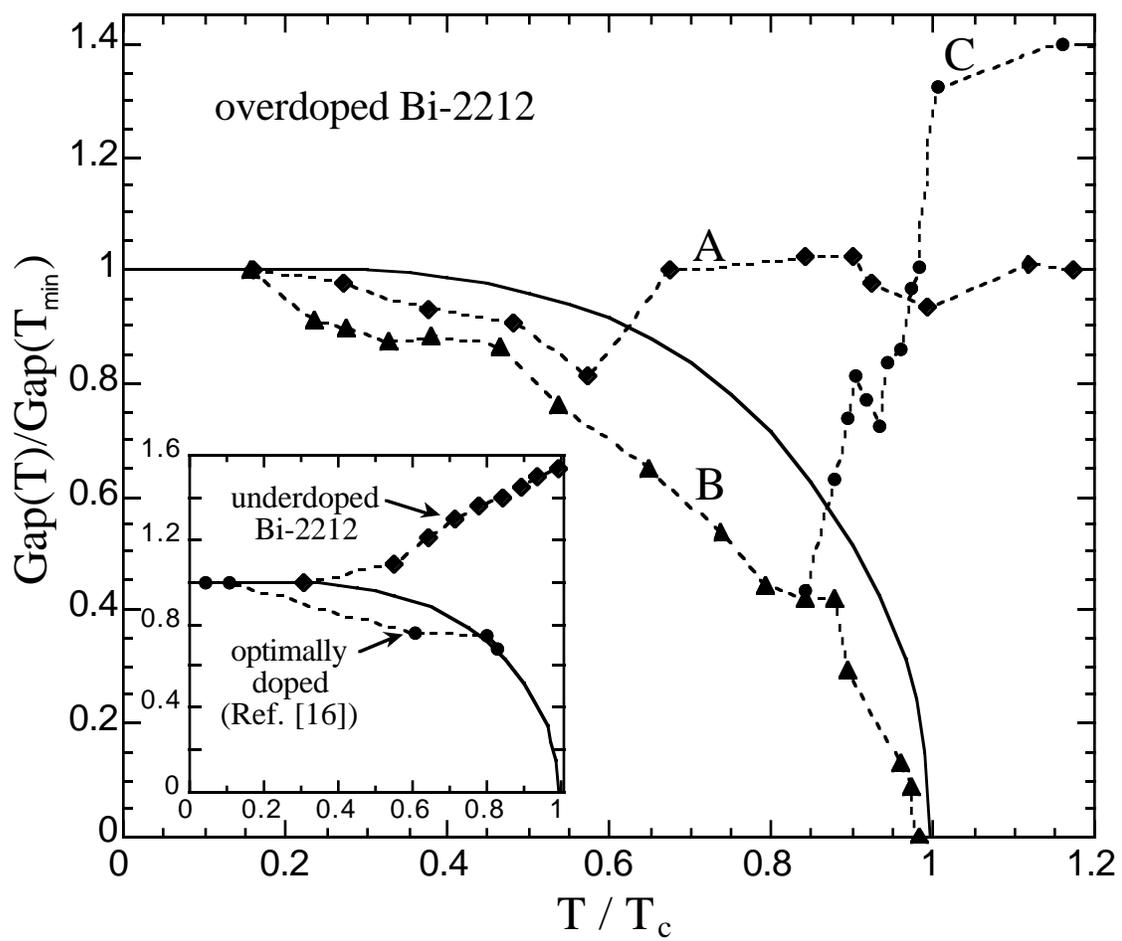

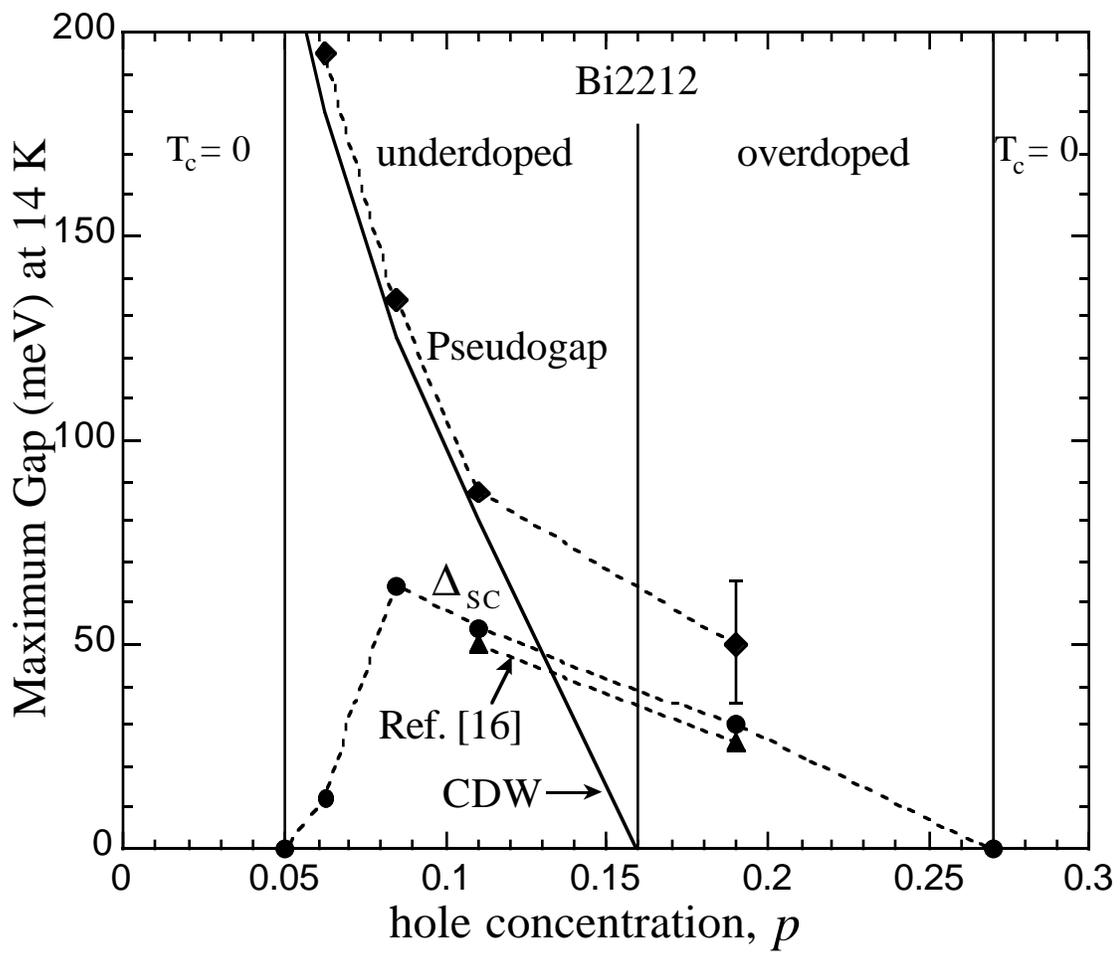